\documentclass[letterpaper, 10 pt, conference]{ieeeconf}

\IEEEoverridecommandlockouts

\usepackage[utf8]{inputenc}
\usepackage[T1]{fontenc}
\usepackage{amssymb} 
\usepackage{amsmath} 
\usepackage{array} 
\usepackage{graphicx}
\newcolumntype{L}[1]{>{\raggedright\let\newline\\\arraybackslash\hspace{0pt}}m{#1}}
\newcolumntype{C}[1]{>{\centering\let\newline\\\arraybackslash\hspace{0pt}}m{#1}}
\newcolumntype{R}[1]{>{\raggedleft\let\newline\\\arraybackslash\hspace{0pt}}m{#1}}
\usepackage{caption} 
\usepackage{makecell}
\usepackage{tabularray}
\usepackage{changepage}
\usepackage{multirow}

\title{\LARGE \bf
Synergistic Formulaic Alpha Generation for \\
Quantitative Trading based on Reinforcement
Learning
}

\author{Hong-Gi Shin, Sukhyun Jeong, Eui-Yeon Kim, Sungho Hong, Young-Jin Cho, and Yong-Hoon Choi \\
Division of Robotics, Kwangwoon University, Seoul 01897, South Korea \\
Email: \{ghdrl95, jayze3736, dmldussla93, gyogook608, yjaycho\}@gmail.com, yhchoi@kw.ac.kr
}

\begin{document}

\maketitle
\thispagestyle{empty}
\pagestyle{empty}

\begin{abstract}

Mining of formulaic alpha factors refers to the
process of discovering and developing specific factors or indicators (referred to as alpha factors) for quantitative trading in
stock market. To efficiently discover alpha factors in vast search
space, reinforcement learning (RL) is commonly employed. This
paper proposes a method to enhance existing alpha factor mining approaches by expanding a search space and utilizing pretrained formulaic alpha set as initial seed values to generate synergistic formulaic alpha. We employ information coefficient (IC)
and rank information coefficient (Rank IC) as performance
evaluation metrics for the model. Using CSI300 market data, we
conducted real investment simulations and observed significant
performance improvement compared to existing techniques. \\


\end{abstract}

\section{INTRODUCTION}

Artificial intelligence plays a vital role in enhancing
profitability in stock investment and the financial sector.
Specifically, quantitative trading involves an automated
approach to buying and selling stocks to maximize the final
net asset value. Machine learning models trained on technical
stock trading data (e.g., open, close, high, low, volume, etc.)
actively participate in the trading process. Research on
quantitative trading utilizing reinforcement learning considers
the volatility and noise in stock data, addressing gaps between
signal-based trading decisions. Reinforcement learning (RL)
constructs end-to-end models, bypassing issues related to
prediction. This approach proves suitable for various
quantitative trading tasks such as algorithmic trading and
portfolio management [1]. However, this field still faces
several limitations. For instance, given the importance of risk
management, there is a pressing need for explainable AI due
to the black-box nature of agents [2].
\\

Formulaic alpha factor mining, a subset of artificial
intelligence research in the stock market, involves the process
of generating formulas with high correlations to future returns
from raw features associated with stock trading. Typically, it
calculates correlations between alpha factor values and future
returns and identifies factors showing high correlations as
'alpha factors,' believed to bring returns exceeding the market
performance. Symbolic factors are equated with formulaic
factors and are typically expressions generated using various
operators and operands. Traditionally, human intervention
was involved in expression creation, but recently, machine
learning models are used to automatically generate them.
When conducting symbolic factor mining based on machine
learning, it requires a search space that specifies which
operators and operands to use in creating symbolic factors, as
well as a search algorithm to find the optimal symbolic factors.
\\

There have been studies using genetic algorithms to find a
single formulaic alpha factor [3-7]. Genetic algorithms aim to
start from various initial factors and use an evolutionary
mechanism to generate the optimal factor. However,
explaining the complex stock market with a single alpha factor
is challenging, so it is common to combine alpha factors with
low correlations among them. Existing methods for alpha
factor generation have prioritized the performance of
individual alpha factors without considering the performance
of combined alpha factors. Therefore, there are limitations in
finding a set of alpha factors that synergistically contribute to
each other.
\\

To address these limitations, Yu \textit{et al}. [8] proposed a new
framework aimed at optimizing the performance of alpha
factor combination during formulaic alpha factor mining. In
their study, they explored the space for alpha factor generation
using proximal policy optimization (PPO)-based
reinforcement learning [9] and updated weights of generated
alpha factor set through gradient descent. The framework
proposed in the paper demonstrated higher correlations with
future returns compared to a single alpha factor generated by
a genetic algorithm. However, unlike other studies on alpha
factor generation, the limited number of operators in the search
space restricts the ability to generate a wide variety of alpha
combinations. These limitations highlight the need to expand
the search space. But excessively expanding the search space
raises complex issues, such as requiring a more sophisticated
policy architecture and a more efficient search algorithm.
\\

In this paper, we build upon the research of Yu \textit{et al}. [8]
by proposing an enhanced initialization method that defines a
more extensive search space and initializes it with pregenerated seed formulaic alpha set, thereby leveraging the
strengths of RL-based search algorithms. Our approach
improves the performance compared with previous synergistic
formulaic alpha factor methodologies. To assess performance
against previous researches, we use CSI300 market data for
the same period. We sequentially apply the proposed
technique, observe improvements in performance, and
evaluate investment outcomes through investment simulations
in comparison with previous methods.
\\

The structure of this paper is as follows. In Section II, we
present the improvements compared to previous synergistic
formulaic alpha factor mining techniques. Section III provides
a detailed explanation of data collection and processing
methods, experimental settings, and results. Finally, in Section
IV, we draw conclusions from this research and discuss
directions for future studies.

\section{PROPOSED METHOD}
We adopt the alpha definition employed in [8]. We trade
for a period of \textit{T} days and consider \textit{n} stocks in the stock
market. For each trading day $\textit{t} \in \{1, \ldots, \textit{T}\}$
each stock \textit{i} corresponds to a feature vector $\textbf{x}_{i, t} \in \mathbb{R}^{mt}$
where, \textit{m} is the
number of raw features such as opening and closing prices.

\begin{table}[t]
\centering
\captionsetup{font=footnotesize}
\captionsetup{justification=centering}
\caption{Information on the tokens used in the experiment. \\
Tokens that were added are indicated in bold.}
\begin{tabular}{l|l}
\Xhline{3\arrayrulewidth}
\multicolumn{1}{c|}{\textbf{Category}}       & 
\multicolumn{1}{c}{\textbf{Symbols}}
\\ \Xhline{3\arrayrulewidth}

Features                & open, close, high, low, volume, VWAP  \\

\hline
\multirow{5}{*}{Operator} 

                        & Abs, Log, Add, Sub, Mul, Div, Greater, Less, \\
                        & Ref, Mean, Std, Var, Sum, Max, Min, Med, \\
                        & Mad, Delta, WMA, EMA, Cov, Corr, \\
                        & \textbf{Sign, CSRank, Product, Scale, Pow, Skew,} \\
                        & \textbf{Kurt, Rank, Delta, Argmax, Argmin, Cond} \\
\hline         
Times deltas            & \textbf{5}, 10, 20, 30, 40, 50, \textbf{60, 120, 252}   \\
\hline
\multirow{2}{*}{Constant}               
                        & -30.0, -10.0, -5.0, -2.0, -1.0, -0.5, -0.01, 0.5,\\
                        &  1.0, 2.0, 5.0, 10.0, 30.0 \\
\hline
Sequence                & BEG(begin), SEP(end of expression)   \\
indicators  \\   
\hline

\end{tabular}
\end{table}
\noindent
With the feature vectors of all stocks on a trading day $\textbf{X} \in \mathbb{R}^{n \times mt}$ consisting of $n$ feature vectors, the alpha factor $f$ is defined as a mapping function that converts the feature vectors $\textbf{X} \in \mathbb{R}^{n \times mt}$ to alpha values $f(\textbf{X}) \in \mathbb{R}^n$. Finally, alpha values $\textbf{z}$ are obtained as $\textbf{z} = \sum_{j=1}^{k} w_j f_j(\textbf{X})$ where $k$ is the number of elements within the alpha set $\mathcal{F} = \{f_1, \ldots, f_k\}$ and their weights $\mathcal{W} = \{w_1, \ldots, w_k\}$. To calculate the correlation between the alpha values and the real stock trend $\textbf{y} \in \mathbb{R}^n$, we used IC as an metric. Also, we based on [8] as a method for generating alpha sets that can maximize IC. The alpha mining system of [8] consists of an alpha generator that produces alpha factors and a combination model that optimizes $\mathcal{W}$ to maximize the IC of the alpha set over the training data.
\\

We propose a search space expansion of the alpha
generator for diversity of alpha factors and initialization with
seed alphas to effectively navigate a large search space.

\mbox{}
\\
\subsection{Expanding Search Space}
This paper utilizes reinforcement learning to generate a
wide range of formulaic alpha factors by exploring a much
broader search space, thus ensuring the creation of diverse
factors. Table 1 depicts the search space used in the paper. We
have incorporated new operators presented in [10] to those
used in previous research [8], while excluding industry
classification data. Additionally, to gain a deeper
understanding of trend changes, we have included a constant
range of \{5, 60, 120, 252\}, allowing for the reflection of both
short-term and long-term trends. Table 2 provides detailed
descriptions of the operators added in this paper.

\begin{table}[t]
\centering
\captionsetup{font=footnotesize} 
\captionsetup{justification=centering} 
\caption{Description of the operators added in the paper.}
\begin{tabular}{c|l}
\Xhline{3\arrayrulewidth}
\multicolumn{1}{c|}{\textbf{Operator}}       & 
\multicolumn{1}{c}{\textbf{Description}}
\\ \Xhline{3\arrayrulewidth}
\multirow{2}{*}{Sign($x$)}               
                                & Returns 0 if the given \textit{x} value is 0, 1 if it is\\
                                & positive, and -1 if it is negative.  \\
\hline
\multirow{4}{*}{CSRank(\textit{x})}         
                        & The cross-sectional rank (CSRank) is an \\
                        & operator that returns the rank of the current \\
                        & stock's feature value \textit{x} relative to the feature \\
                        & values of all stocks on today's date. \\
\hline

\multirow{4}{*}{Product(\textit{x, t})}    
                        & It returns the product of the feature values for \\
                        & each date from the current date up to \textit{t} days \\
                        & ago. \\
                        & \text{Product}(\textit{x, t}) = $\prod_{i=0}^{t} \textit{x}_{t-i}$ \\
                        
\hline         
\multirow{4}{*}{Scale(\textit{x})}   
                        & It returns the value obtained by dividing the \\
                        & current feature value \textit{x} by the total sum of the \\
                        & absolute values of the feature. \\
                        & \text{Scale}(\textit{x}) = $\frac{\textit{x}}{\sum_i |\textit{x}_i|}$ \\
\hline
Pow(x,y)                & Pow(\textit{x, y}) = $\textit{x}^\textit{y}$   \\
\hline
\multirow{6}{*}{Skew(\textit{x})}    
                        & Skewness. It represents the asymmetry of a \\
                        & data distribution and is expressed using the \\
                        & third standard moment. \\
                        & $\mu = E(\textit{x}), \quad \mu_i = E[(\textit{x} - \mu)^i],$ \\
                        & $\text{Skew}(\textit{x}) = \left( \frac{\mu_3}{\mu_2^{1.5}} \right) = \left( \frac{E[(\textit{x} - \mu)^3]}{(E[(\textit{x} - \mu)^2])^{1.5}} \right)$ \\
\hline
\multirow{5}{*}{Kurt(\textit{x})}                
                        & Kurtosis, a value indicating the peakedness of \\
                        & a data distribution, represents how much the \\
                        & observations are clustered around the mean. \\
                        & \text{Kurt}(\textit{x}) = $\frac{\mu_4}{\mu_2^2} - 3 = \left( \frac{E[(\textit{x} - \mu)^4]}{(E[(\textit{x} - \mu)^2])^2} \right) - 3$ \\
\hline
\multirow{4}{*}{Rank(\textit{x, t})}                
                        & Time-series rank (Rank), an operator that \\
                        & returns the rank of the current feature value \textit{x} \\
                        & among feature values from the current date up \\
                        & to \textit{t} days ago. \\
\hline
\multirow{4}{*}{Delta(\textit{x, t})}              
                        & An operator that returns the difference \\
                        & between the current feature value \textit{x} and the \\
                        & feature value from \textit{t} days ago.   \\
                        & $\text{Delta}(\textit{x, t}) = \textit{x} - \text{Ref}(\textit{x, t})$ \\
\hline
\multirow{3}{*}{Argmax(\textit{x, t})}             
                        & An operator that returns the date when the \\
                        & feature value \textit{x} was the highest within the \\
                        & period from the current date up to \textit{t} days ago. \\
\hline
\multirow{3}{*}{Argmin(\textit{x, t})}              
                        & An operator that returns the date when the \\
                        & feature value \textit{x} was the lowest within the \\ 
                        & period from the current date up to \textit{t} days ago. \\
\hline
\multirow{2}{*}{Cond(\textit{x,y,t,f})}       
                        & An operator that returns \textit{t} if $\textit{x} > \textit{y}$ is true, \\ 
                        & and \textit{f} if it is false. \\ 
\hline

\end{tabular}
\end{table}

\subsection{Initialization with Seed Alphas}
The policy must navigate the given search space to
generate \textit{k} alpha factors. This implies that as we expand the
search space, the complexity of the search space that the policy
must explore to generate optimal alpha factors increases,
necessitating more sophisticated architectural structures for
the policy and refinement of the search algorithm. This study
sets pre-generated formulaic alpha set as the initial seed alpha
set and then performs alpha set mining based on it. By storing
a combination of pre-generated superior alpha set in the replay
buffer, it is possible to bias the search space that the policy
needs to explore, thereby reducing the complexity of the
search and enabling the creation of synergistic alphas in fewer
steps. The process is carried out in two stages. In the first stage,
if there are no previously known alpha factors, the alpha set is
initialized as empty(w/o initial seed alpha factor).
Subsequently, mega alphas are generated through synergistic
formulaic alpha mining. In the second stage, if there are
already known (or generated) alpha factors, the alpha set is
initialized with these factors included. Then, by continuing to
generate mega alphas in a complementary manner, a more
enhanced predictive model is ultimately constructed.

\begin{table}[t!]
\captionsetup{font=footnotesize}
\captionsetup{justification=centering}
\caption{Main results on CSI 300. Values outside parentheses are the means, and values inside parentheses are the standard deviations across 5 runs.}
\centering
\begin{tabular}{l|c|c}
\Xhline{3\arrayrulewidth}
\multirow{2}{*}{Method} & \multicolumn{2}{c}{CSI 300} \\
\cline{2-3}
 & IC($\uparrow$) & Rank IC($\uparrow$) \\
\Xhline{3\arrayrulewidth}
\multirow{2}{*}{Baseline [8]} & 0.045 & 0.058 \\
 & (0.0036) & (0.0006) \\
\hline
\multirow{2}{*}{Ours (Expanding search space)} & 0.069 & 0.073 \\
 & (0.0079) & (0.010) \\
\hline
Ours (Expanding search space + & 0.071 & 0.071 \\
Initialization with 101 alpha) & (0.0086) & (0.008) \\
\hline
Ours (Expanding search space + & 0.085 & 0.087 \\
Initialization with generated alpha set) & (0.003) & (0.003) \\
\hline
\end{tabular}
\end{table}

\begin{figure}[t!]
    \centerline{\includegraphics[width = 7.5cm]{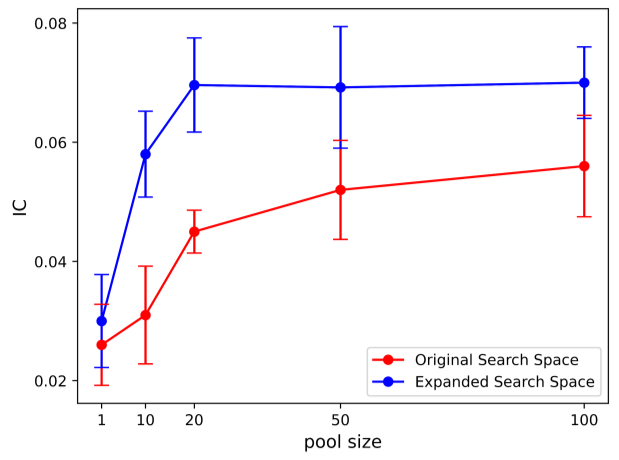}}
    \captionsetup{font=footnotesize}
    \caption{ Comparison of IC changes according to pool size variations 
in the Combination model on CSI300. Display average IC after training 
five random seeds for each pool size. Red: Original search space, Blue: 
Expanded search space.}

\end{figure}

\section{EXPERIMENTS}
\subsection{Experiment Environment}
The data used in this study is from the Chinese A-shares
market, which includes six features: Open, Close, High, Low,
Volume, and volume weighted average price (VWAP). To
prevent survivorship bias, the listing date of each stock was
used as its index inclusion date. Since this study only considers long positions, the dates when stocks are excluded from the
index were not considered. 
The target variable is the stock
price change percentage in 20 days later. 
The training set spans from January 1, 2009, to December 31, 2018, the validation set from January 1, 2019, to December 31, 2019, and the test set from January 1, 2020, to December 31, 2021. For the backtest framework, Qlib [11] was utilized.
\\

To evaluate the performance of the proposed technique,
we sequentially applied an expanded search space and
initialization with the generated alpha set, setting the model's combination size to 20. Additionally, to verify accuracy, five
different random seeds were applied. As a performance metric,
we use the Pearson correlation coefficient to measure the
correlation between the target variable and alpha factor.
Additionally, we use the Spearman rank correlation
coefficient to measure rank-based correlations.

\begin{table}[t]
\captionsetup{font=footnotesize}
\captionsetup{justification=centering}
\caption{Top 5 formulas from Alpha 101 [10] with the highest IC.}
\centering
\begin{tabular}{l|@{}l|l}
\Xhline{3\arrayrulewidth}
\multirow{2}{*}{Alpha\#}   &  \multicolumn{1}{l|}{\multirow{2}{*}{Expression}} & IC in \\
{}      &  {} & test set\\
\Xhline{3\arrayrulewidth}
Alpha    &  \multirow{2}{*}{(-1 * \text{Corr}(\text{open}, \text{volume}, 10))} & \multirow{2}{*}{0.035} \\
006      &  {} & {} \\
\hline
\multirow{4}{*}{\shortstack[l]{Alpha \\ 099}}         &  (Less(CSRank(Corr(Sum(((high + low) /   & \multirow{4}{*}{0.032} \\
{}      &  2), 19.8975), Sum(Mean(volume, 60),  & {} \\
{}        &  19.8975), 8.8136)), CSRank(Corr(low, & {} \\
{}         &  volume, 6.28259))) * -1) & {} \\
\hline
\multirow{3}{*}{\shortstack[l]{Alpha \\ 061}}         &  Less(CSRank((vwap - Min(vwap,  & \multirow{3}{*}{0.031} \\
{}      &  16.1219))), CSRank(Corr(vwap, & {} \\
{}        &  Mean(volume, 180), 17.9282))) & {} \\
\hline
\multirow{3}{*}{\shortstack[l]{Alpha \\ 014}}         &  ((-1 * CSRank(Delta(Div(Sub(close,   & \multirow{3}{*}{0.028} \\
{}      &  Ref(close, 1)), close), 3))) * Corr(open,   & {} \\
{}        &  volume, 10)) & {} \\
\hline
\multirow{4}{*}{\shortstack[l]{Alpha \\ 035}}         &  ((Rank(volume, 32) * (1 - Rank(((close +  & \multirow{4}{*}{0.024} \\
{}      &  {high) - low), 16))) * (1 -} & {} \\
{}        &  Rank(Div(Sub(close, Ref(close, 1)), & {} \\
{}         &  close), 32))) & {} \\
\Xhline{3\arrayrulewidth}
\end{tabular}
\end{table}

\begin{figure}[t!]
    \centerline{\includegraphics[width = 7.5cm]{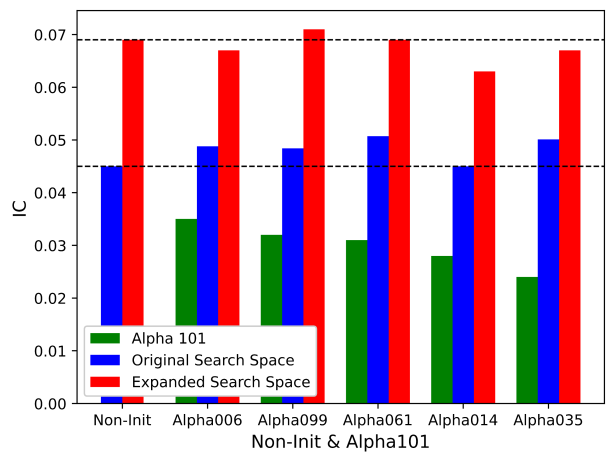}}
    \captionsetup{font=footnotesize}
    \caption{IC change during the Test period when initializing the Alpha 
set with Alpha 101's formulaic alpha factor. Non-Init: Training without 
initializing with a separate formulaic alpha. Blue: Original search 
space, Red: Expanded search space. Green: IC for the test set of the 
existing Alpha 101 formula.}
\end{figure}

\subsection{Main Result}
Table 3 presents the results of the performance comparison
with [8]. Upon examining the experimental results, we
observed that the performance differences among all models 
were minimal with changes in seed value. However,
performance improvements were observed when expanding
the search space and sequentially implementing the alpha set
initialization strategy. While [8] showed a significant value
difference between IC and RankIC, the proposed technique
maintained value differences within the range of standard
deviation.
\\

\begin{figure}[t!]
    \centerline{\includegraphics[width = 7.5cm]{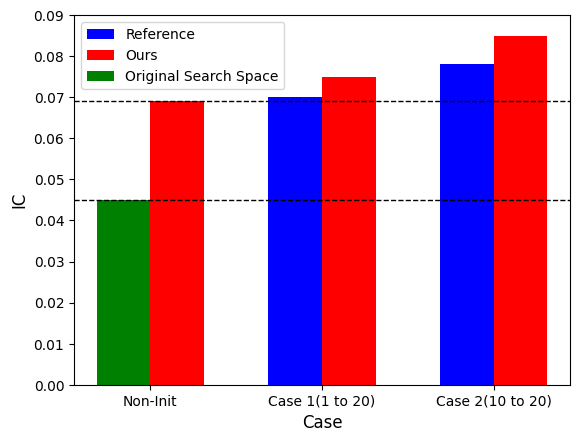}}
    \captionsetup{font=footnotesize}
    \caption{IC change during the Test period when initializing the Alpha 
set with the created alpha set. Non-Init: Training without inserting any 
separate formulaic alpha. Blue: Original search space, Red: Expanded 
search space. Green: IC for the test set of alphas created when the pool 
size is 10.}
\end{figure}

\begin{table}[t]
\captionsetup{font=footnotesize}
\captionsetup{justification=centering}
\caption{Parameters used in backtesting.}
\centering
\begin{tabular}{l|r}
\hline
\Xhline{3\arrayrulewidth}
\textbf{Parameters} &  \multicolumn{1}{l}{\textbf{Value}} \\
\Xhline{3\arrayrulewidth}
\hline
Top K & 50 \\
\hline
Swap N & 5 \\
\hline
Minimum number of holding days \( H \) & 20 days \\
\hline
Top K enter threshold \( E_{th} \) & 0.0 \\
\hline
Backtest dates & 2020-01-01 - 2021-12-31 \\
\hline
Survivorship bias & Not included \\
\hline
Backtest platform & Qlib [11] \\
\hline
PPO agent seed & 0, 1, 2, 3, 4 \\
\hline
Size of alpha pool & 20 \\
\hline
\end{tabular}
\end{table}

\subsection{Case Study 1: Expanding Search Space}
In this paper, we conducted experiments to investigate the
impact of the expansion of operators and operands on
performance. The experiment observed changes in IC by
varying the pool sizes to 1, 10, 20, 50, and 100.
\\

Figure 1 shows the results of IC changes for each pool size
based on the CSI300 data. These results demonstrate a general
trend of performance improvement with increasing pool sizes
in [8]. However, the expanded search space proposed in our
study showed that even with smaller pool sizes, IC was higher
for the test period. In the expanded search space, we observed
performance improvement up to a pool size of 20, after which
no further changes in performance were noted.

\subsection{Case Study 2: Initialization with alpha 101}
In this paper, we conducted experiments to determine
whether initializing with [10]'s alpha factors impacts the
learning effectiveness. During the training process, we
selected 5 formulaic alpha factors from the top 101 alphas with 
high IC for the CSI300 index, excluding alphas that use
'indneutralize' and 'market cap'. The formulas used are
presented in Table 4. The pool size for the combination model
was set to 20, ensuring that the formulas proposed in [10] were
included in the initial alpha set. At the same time, the setup
allowed for the possibility of these alpha factors being
removed from the alpha set by the combination model during
training. For the comparison, we named the original method
without [10]'s seed alpha factors as 'Non-Init' and observed the
performance differences depending on the presence or absence
of [10]'s seed alpha factors. We also observed the performance
changes in both the original search space and the expanded
search space. In this experiment, training was conducted with
five different random seeds for the CSI300 index.
\\

Figure 2 presents the results when the alpha set was
initialized using the formulas proposed in [10]. In the original
search space, the average IC increased by 0.0034, indicating a
performance difference within the standard deviation.
Conversely, in the expanded search space, the average
decreased by 0.0016, showing no significant changes in
performance in both search spaces. The formulas of alpha 101
used in the experiment displayed low ICs during the test set
period. In the original search space, only alpha 099 was
included in the alpha set for one of the five random seeds,
while the other formulaic alpha factors from alpha 101 were
dropped during the generation process. These results suggest that the formulaic alpha factors of alpha 101 were generally
removed from the alpha set creation process due to their
overall low performance.

\begin{figure}[t!]
    \centerline{\includegraphics[width = 9cm]{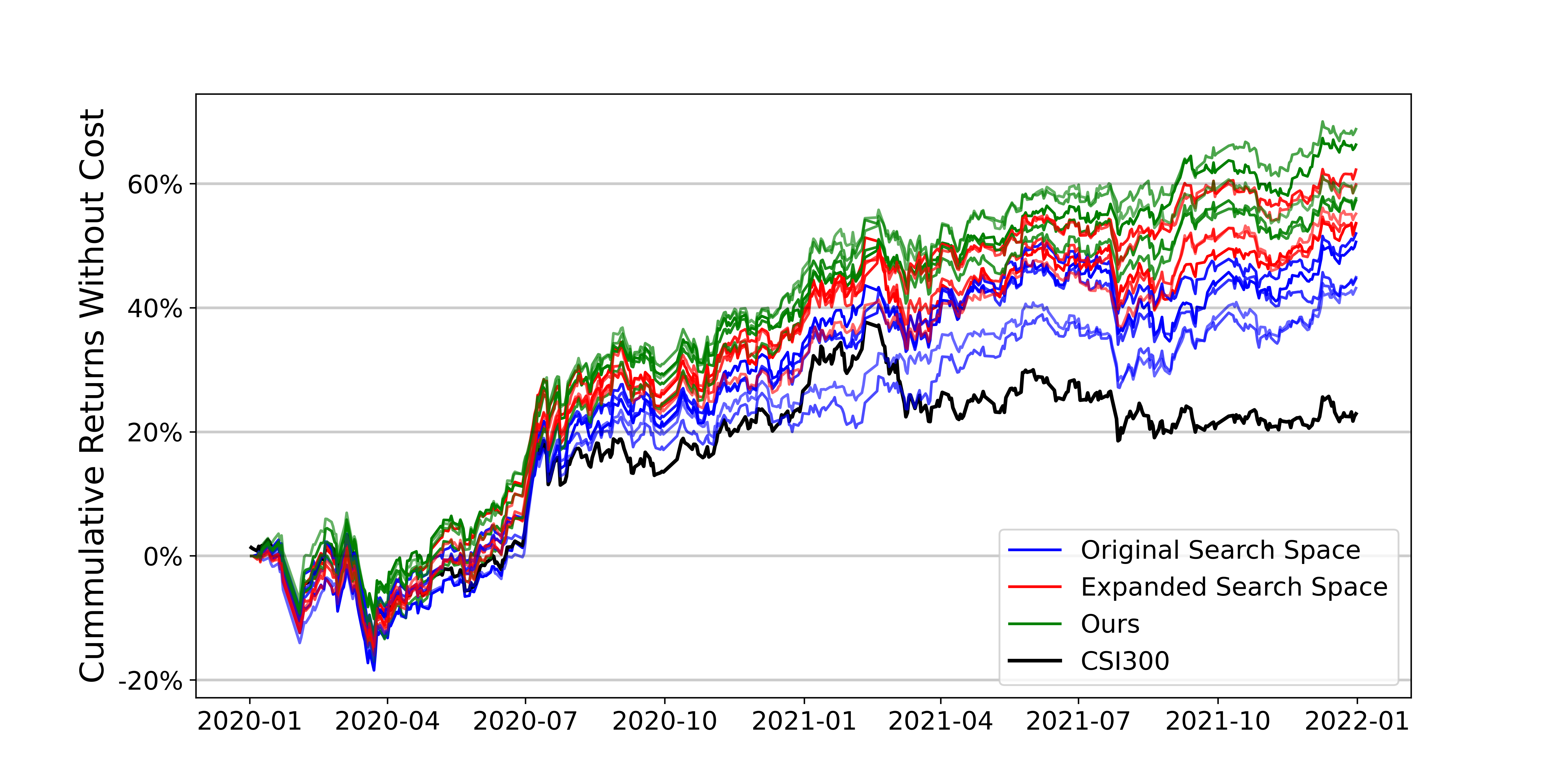}}
    \captionsetup{font=footnotesize}
    \caption{Cumulative return changes during the test period with the 
created mega alpha set. Set pool size to 20 and display individually after 
training five random seeds. Black: CSI300 index for the test set period. 
Blue: alpha set created with original search space, red: alpha set created 
with expanded search space, green: alpha set created after initializing with 
alpha set generated when pool size is 10.}
\end{figure}

\subsection{Case Study 3: Initialization with pre-generated alpha set}  
This experiment was designed to determine whether the
initialization of alpha set using synergistic formulaic alpha impacts the learning effectiveness. During the expanded search
space experiment, we measured the test ICs of the alpha set
generated when the combination model's pool size was 10. To
conduct a comparative experiment with alpha 101, we selected
a single formulaic alpha factor from the alpha set that showed
the highest IC in the test set for use in alpha set initialization.
The selected alpha set demonstrated an IC of 0.078 in the test
set, and from this, a single formulaic alpha factor with an IC
of 0.07 was chosen.
\\

The pool size of the combination model was set to 20, and
5 different random seeds were used for training and evaluation
of each technique. During the training, selected formulaic alpha factors were added to the alpha set, and if necessary, could
be removed by the combination model. This experiment was
conducted on the CSI300 index, and we observed the impact
of initialization with seed alphas from the experiment by comparing performances based on whether the empty alpha set,
marked as 'Non-Init,' were initialized or not.
\\

Figure 3 shows the experimental results when alpha set
with a pool size of 20 was initialized using alpha set with a
pool size of 10. Initializing with a single formulaic alpha factor
resulted in an increase of 0.006 in IC compared to initializing
with empty alpha set with a pool size of 20, and an increase of
0.005 compared to a pool size of 10. When initialized with 10
formulaic alpha factors, there was an increase of 0.016 in IC
compared to initializing with empty alpha set with a pool size
of 20, and an increase of 0.007 compared to a pool size of 10.
The significant IC differences observed in experiments with
the same pool size demonstrate that the method of initialization with the initial seed alpha set can affect model performance.

\subsection{Backtesting}  
In this paper, the backtesting environment for the proposed
investment strategy utilized a \textit{Top-K/Swap-N} based long only
strategy. \textit{Top-K/Swap-N} involves selecting the top K stocks
based on the highest alpha values to form the portfolio. Daily,
at the close of the stock market, the alpha values of stocks in
the portfolio are compared with those not included in the portfolio. Up to N stocks are then sold, and up to N new stocks
are purchased based on this comparison. Additionally, stocks
initially purchased must be held for at least a minimum
number of days (\textit{H}) before they can be sold. A \textit{Top-K Enter
Threshold($E_{th}$)} 
is also set, where the value must be higher
than this threshold at the time of purchase and lower at the
time of sale. Table 5 presents the parameters used for
backtesting.
\\

Figure 4 displays the cumulative returns after backtesting
for each technique trained with 5 random seeds. It is observed
that the proposed technique consistently recorded higher
cumulative returns across all seeds compared to the existing
techniques. While some seeds in the existing methods failed
to achieve excess returns over the CSI300 index during certain
periods, the application of the proposed technique resulted in
excess returns over the index for all seeds.

\section{CONCLUSION}
In this paper, we explored the efficiency of using
reinforcement learning to generate synergistic formulaic alpha
collections, and confirmed the potential of reinforcement
learning in creating formulaic alpha factors from a vast search
space, demonstrating its capability to produce alpha set with
high IC. We found that initializing with pre-generated
formulaic alphas led to the creation of superior performing
alpha set. Additionally, we expanded the search space to
integrate various operators and operands, and confirmed that
this expansion contributed to improved results. However,
there are limitations due to the complexity arising from the
formula length restriction inherent in the model structure. To
address this, we proposed the use of predefined auxiliary
indicators. We also identified a problem where IC deteriorated
at the beginning of the training process due to not resetting the
experience buffer, suggesting the need to simultaneously
initialize the alpha set and experience buffer in future to
resolve this issue.
\\

Finally, we proposed the necessity of observing
performance in other markets. While this paper conducted
experiments using the CSI300, further experiments in various
financial markets like the CSI500, NASDAQ, and KOSPI are
planned to additionally verify the universality of our approach.

\addtolength{\textheight}{-12cm}

\section*{ACKNOWLEDGMENT}
This work was supported in part by the National Research
Foundation of Korea (NRF) Grant funded by the Korea
Government MSIT under Grant 2021R1F1A1064080.

\end{document}